\documentstyle[12pt]{article}
\catcode`\@=11
\def\section{\@startsection {section}{1}{0pt}{-3.5ex plus -1ex minus
 -.2ex}{2.3ex plus .2ex}{\raggedright\large\bf}}
\catcode`\@=12
\newskip\humongous \humongous=0pt plus 1000pt minus 1000pt

\newif\ifdtup

\def\oldreffmt#1{\rlap{[#1]} \hbox to 2\parindent{}}

\def\figfmt#1{\rlap{Figure {#1}} \hbox to 1in{}}

\def\beq{\begin{equation}}
\def\eeq{\end{equation}}
\def\bea{\begin{eqnarray}}

\def\eea{\end{eqnarray}}
\catcode`\@=11
\def\eqnarray{\stepcounter{equation}\let\@currentlabel=\theequation
\global\@eqnswtrue
\global\@eqcnt\z@\tabskip\@centering\let\\=\@eqncr
\gdef\@@fix{}\def\eqno##1{\gdef\@@fix{##1}}%
$$\halign to \displaywidth\bgroup\@eqnsel\hskip\@centering
  $\displaystyle\tabskip\z@{##}$&\global\@eqcnt\@ne
  \hskip 2\arraycolsep \hfil${##}$\hfil
  &\global\@eqcnt\tw@ \hskip 2\arraycolsep $\displaystyle\tabskip\z@{##}$\hfil
   \tabskip\@centering&\llap{##}\tabskip\z@\cr}

\def\@@eqncr{\let\@tempa\relax
    \ifcase\@eqcnt \def\@tempa{& & &}\or \def\@tempa{& &}
      \else \def\@tempa{&}\fi
     \@tempa \if@eqnsw\@eqnnum\stepcounter{equation}\else\@@fix\gdef\@@fix{}\fi
     \global\@eqnswtrue\global\@eqcnt\z@\cr}

\newcount\hour
\newcount\minute
\newtoks\amorpm
\hour=\time\divide\hour by 60
\minute=\time{\multiply\hour by 60 \global\advance\minute by-\hour}
\edef\standardtime{{\ifnum\hour<12 \global\amorpm={am}%
	\else\global\amorpm={pm}\advance\hour by-12 \fi
	\ifnum\hour=0 \hour=12 \fi
	\number\hour:\ifnum\minute<10 0\fi\number\minute\the\amorpm}}
\edef\militarytime{\number\hour:\ifnum\minute<10 0\fi\number\minute}

\def\draftlabel#1{{\@bsphack\if@filesw {\let\thepage\relax
   \xdef\@gtempa{\write\@auxout{\string
      \newlabel{#1}{{\@currentlabel}{\thepage}}}}}\@gtempa
   \if@nobreak \ifvmode\nobreak\fi\fi\fi\@esphack}
	\gdef\@eqnlabel{#1}}
\def\@eqnlabel{}
\def\@vacuum{}
\def\marginnote#1{}
\def\draftmarginnote#1{\marginpar{\raggedright\scriptsize\tt#1}}

\def\draft{\oddsidemargin -.5truein
	\def\@oddhead{\sl \phantom{\today\quad\militarytime} \hfil
	\smash{\Large\sl DRAFT} \hfil \today\quad\militarytime}
	\let\@evenhead\@oddhead
	\let\label=\draftlabel
	\let\marginnote=\draftmarginnote
	\def\ps@empty{\let\@mkboth\@gobbletwo
	\def\@oddfoot{\hfil \smash{\Large\sl DRAFT} \hfil}
	\let\@evenfoot\@oddhead}
   	\def\@eqnnum{(\theequation)\rlap{\kern\marginparsep\tt\@eqnlabel}%
	\global\let\@eqnlabel\@vacuum}  }

\catcode`\@=12

\def\lae{\smash{\,\lower .5 ex \hbox{$\,\stackrel<\sim\,$}}}
\def\gae{\smash{\,\lower .5 ex \hbox{$\,\stackrel>\sim\,$}}}

\def\L{{\cal L}}

\def\beq{\begin{equation}}
\def\eeq{\end{equation}}

\def\sutw{${\rm SU}(2)_W$}

\begin{document}
\begin{titlepage}
\begin{center}
July 9, 1992\hfill    WIS-92/56/JULY-PHYS

\vskip 1 cm

{\large \bf On the Unmixed 17 keV Neutrino}

\vskip 1 cm

Miriam Leurer

\vskip 1 cm

{\em Department of Nuclear Physics\\
The Weizmann Institute\\
Rehovot 76100\\
ISRAEL}

\end{center}

\vskip 1 cm

\begin{abstract}

Recently, it was suggested that the 17 keV neutrino does not mix with the
electron neutrino in the weak interactions. Instead, the $\beta$ decay mode
involving the 17 keV neutrino is induced by a completely new interaction,
presumably mediated by leptoquarks. A previous model for the ``unmixed 17
keV neutrino" suffers from difficulties with experimental data and
cosmological constraints. Here we present an alternative model in which
these difficulties are resolved.
\end{abstract}
\newpage

\end{titlepage}

In a recent paper \cite{MPRT} Masiero, Pakvasa, Roulet and Tata argued that the
kink claimed to be seen in the $\beta$ decay spectrum \cite{simps} is not
necessarily the signal of a 17 keV neutrino mixing at the $1\%$ level with the
electron neutrino. Instead, they suggested that there is a completely new
interaction, mediated by scalar leptoquarks, that induces the decay
$d\longrightarrow ue\nu_x$, where $\nu_x$ is the 17 keV neutrino. The strength
of the new interaction is 10$\%$ of the strength of the standard weak
interaction that induces the usual $d\longrightarrow ue\bar\nu_e$.

The authors of \cite{MPRT} suggested a specific model for the leptoquark
mediated interaction and mentioned a few problems. First, $\pi$ decay to $e\nu$
is enhanced by $1\%$ relative to the standard model value, and this causes a
significant deviation from $e-\mu$ universality. This problem was sharpened
when new experimental results \cite{Britton} on $\pi$ decay were published,
excluding such a deviation from universality at the 3.5$\sigma$ level. In this
paper we will argue that if there are new interactions that enhance
semileptonic decays to $Xe\nu$ at the 1$\%$ level, then there might also be
interactions that will enhance decays to $X\mu\nu$ at a similar level. We will
show a particular example for such extra interactions.

The second problem mentioned in \cite{MPRT} is cosmological. The 17 keV
neutrino there is none of the three standard LH neutrinos, but rather a new RH
neutrino, which is a singlet of weak isospin. The leptoquark mediated
interactions of this new neutrino turn out to be strong enough to keep it in
equilibrium down to the time of nuclear synthesis, contradicting the latest
analysis of primordial nuclear synthesis \cite{Olive} which do not allow
$N_\nu$ to be larger than 3.6. In our alternative model, the 17 keV neutrino is
the standard LH tau neutrino. No new neutrino degree of freedom is therefore
invoked at the time of nuclear synthesis, and the cosmological problem is
resolved.

In addition, no specific mechanism for the Simpson neutrino decay was suggested
in \cite{MPRT} (although a possible Majoron decay mechanism was alluded to).
Here we wish to advocate a decay mechanism that was first put forward in
\cite{BMR}. There it was suggested that, in case neutrinos were Dirac
particles, the Simpson neutrino decay could proceed purely in the RH sector. We
will offer a particular interaction that will induce such decays.

Another potential difficulty with the model of ref. \cite{MPRT} is that the
polarization of
the emitted electron is RH, in contrast to the standard model prediction of a
LH electron. Experimental measurements of the longitudinal polarization of the
electron in Gamow-Teller type $\beta$ decays support the standard prediction
and exclude the model of ref. \cite{MPRT} at the 2$\sigma$ level. In the model
we will
offer here, the electron emitted in $\beta$ decay is always LH, and the
electron polarization problem does not arise at all.

The authors of \cite{MPRT} presented an elegant  chain of arguments which
seemed to necessarily lead them to their model. We start by briefly repeating
the arguments of \cite{MPRT} and pointing out a loophole that allows for, {\it
e.g.}, our model.

The new effective interaction mediating the decay $d\longrightarrow ue\nu_x$
is:
\beq
\L_{eff}=\sum_i\frac{G_i}{\sqrt2}\bar u O_i d ~ \bar e \tilde{O}_i \nu_x
             ~~+~~h.c.\label{lprimi}
\eeq
where for each $i$, $O_i$ and $\tilde O_i$ are operators, built of Dirac
matrices, and having the same Lorentz group properties; they could differ in
their Parity transformations. The coefficients $G_i$ should generically be of
the order of $0.1G_F$.

To account for the observation of the 17 keV neutrino in the various nuclei
(particularly $^{14}C$, $^{63}Ni$, $^{55}Fe$ and $^{71}Ge$), at least one of
the operators $O_i$ must be $P$ (pseudo scalar), $A$ (axial vector) or $T$
(tensor). However, the possibility of $O_i$ being $P$ and having a coefficient
$G_i$ of the order of 0.1$G_F$ is excluded, since $\pi$ decay to $e\nu$ would
be enhanced in this case by almost five orders of magnitude \cite{shanker}.
Therefore, the interaction $\L_{eff}$ should include a term with $O_i=A$ or
$T$  but not a term with $O_i=P$.

Our purpose is to build a model in  which the interaction (\ref{lprimi}) is
induced by leptoquarks. This means that (\ref{lprimi}) is actually a Fierz
transform. According to the above arguments, we wish this Fierz transform
to include a term with $O_i=A$ or $T$ but no term with $O_i=P$. This implies
that $\L_{eff}$ is of the following form:
\beq
\L_{eff}=4\cdot\sum_i\frac{G_i}{\sqrt2}\bar u \gamma_\mu P_i d ~
                                           \bar e \gamma^\mu \tilde{P}_i \nu_x
             ~~+~~h.c.\label{lmed}
\eeq
where $P_i$ and $\tilde{P}_i$ are LH or RH projection operators:
$\frac{1}{2}(1\pm\gamma_5)$. Obviously, there could be four different terms
in the sum in (\ref{lmed}). The one which we find to be most attractive
has $P_i=P_R$ and $\tilde{P}_i=P_L$. The authors of \cite{MPRT}
however, claimed  that such a term is not \sutw{} invariant, and is therefore
excluded. They then proceeded to argue that the only possible term had
$P_i=\tilde{P}_i=P_R$, and were therefore forced into introducing a
nonstandard RH neutrino as the 17 keV neutrino.

Here we wish to point out that the request for \sutw{} invariance of $\L_{eff}$
is too strong. Since \sutw{} is a broken symmetry, effective interactions need
not be invariant under it. \sutw{} just implies that other, related,
interactions exist. Let us therefore assume that the above interaction includes
only one term, with $P_i=P_R$ and $\tilde P_i=P_L$. This in particular means
that the 17 keV neutrino is the LH weakly interacting $\mu$ or $\tau$
neutrino.
Let us further assume that the interaction is actually mediated by a scalar
leptoquark, and Fierz the interaction back to the original,
leptoquark-mediated, representation:
\beq
\L{eff} = 8\frac{G}{\sqrt2}\bar u_R \nu^x_L \bar e_L d_R  ~~+~~h.c.
              \label{leff}
\eeq
This interaction is induced by a charge $(-\frac{2}{3})$ leptoquark that
couples
both to $\bar e_L d_R$ and to $\bar \nu^x_L u_R$. Such a leptoquark must be a
mixture of two $SU(2)_W \times U(1)_Y$ representations: Both representations
are isospin doublets, but the hypercharges are different: ($-\frac{1}{3}$) for
one and ($-\frac{7}{3}$) for the other. We denote the two leptoquark
isodoublets by $S$ and $\tilde{S}$,
\begin{eqnarray}
S&=&(S_{1/3},S_{-2/3})\nonumber\\
\tilde S&=&(\tilde S_{-2/3},\tilde S_{-5/3})\; ,
\label{S}
\end{eqnarray}
where the subscripts on the members of each multiplet indicate their respective
electromagnetic charges.

The \sutw{} invariant interactions of $S$ and $\tilde S$ with the fermions are
\begin{eqnarray}
\L & = & g(S_{-2/3}\bar e_L d_R + S_{1/3}\bar\nu^e_L d_R)\nonumber\\
       & + & \tilde{g}(\tilde{S}_{-2/3}\bar\nu^x_L u_R+
                       \tilde{S}_{-5/3}\bar{l}^x_L u_R)\; ,
\label{Yukawa}
\end{eqnarray}
where $\nu^x_L$ and ${l}^x_L$ constitute the lepton doublet of the $x$'th
generation. As a result of \sutw{} breaking, $S_{-2/3}$ and $\tilde S_{-2/3}$
mix to form the mass eigenstates $\sigma$ and $\tilde\sigma$:
\begin{eqnarray}
\sigma &=& \cos\theta S_{-2/3}-\sin\theta \tilde{S}_{-2/3}\nonumber\\
\tilde\sigma &=& \sin\theta S_{-2/3}+\cos\theta \tilde{S}_{-2/3}\; .\label{mix}
\end{eqnarray}
The full effective low energy interaction induced by
$\sigma$, $\tilde\sigma$, $S_{1/3}$ and $\tilde S_{-5/3}$ is:
\begin{eqnarray}
\L_{eff} &=&-    4\frac{G^C}{\sqrt2}
                 \left[\bar u_R \gamma^\mu d_R ~ \bar e_L\gamma_\mu \nu^x_L
                     ~~+~~h.c.\right]\nonumber\\
             &&- 4\frac{G^N_l}{\sqrt2}
                 \bar d_R \gamma^\mu d_R~\bar e_L\gamma_\mu e_L
              -  4\frac{G^N_\nu}{\sqrt2}
                 \bar d_R\gamma^\mu d_R~\bar \nu^e_L\gamma_\mu \nu^e_L
                 \nonumber\\
             &&- 4\frac{\tilde G^N_\nu}{\sqrt{2}}
                 \bar u_R\gamma^\mu u_R \bar \nu^x_L \gamma_\mu \nu^x_L
              -  4\frac{\tilde G^N_l}{\sqrt2}
                 \bar u_R\gamma^\mu u_R \bar l^x_L \gamma_\mu l^x_L\; ,
\label{fulll}
\end{eqnarray}
where $G$ and $\tilde{G}$ are the effective four-Fermi couplings,
with superscript $C$ or $N$ to indicate charged currents or neutral currents
processes respectively, and subscript $l$ or $\nu$ to indicate whether charged
leptons or neutrinos are involved. The $G$ and $\tilde G$ coefficients are
given by:
\begin{eqnarray}
\frac{G^C}{\sqrt2}&=&\frac{1}{8}g\tilde{g}\cos\theta\sin\theta
                 \left(\frac{1}{\tilde{M}^2}-\frac{1}{M^2}\right)
\nonumber\\
\frac{G^N_l}{\sqrt2}&=&\frac{1}{8}g^2
                 \left(\frac{\cos^2\theta}{M^2}+
                                     \frac{\sin^2\theta} {\tilde{M}^2}\right)
\nonumber\\
\frac{G^N_\nu}{\sqrt2}&=&\frac{1}{8}\frac{g^2}{M^2_{1/3}}
\nonumber\\
\frac{\tilde G^N_\nu}{\sqrt2}&=&\frac{1}{8}\tilde{g}^2
                 \left(\frac{\sin^2\theta}{M^2}+
                                    \frac{\cos^2\theta}{\tilde{M}^2}\right)
\nonumber\\
\frac{\tilde G^N_l}{\sqrt2}&=&
                 \frac{1}{8}\frac{\tilde{g}^2}{M^2_{-5/3}}
\;,\label{G}
\end{eqnarray}
where $M$, $\tilde{M}$, $M_{1/3}$ and $M_{-5/3}$ are the masses of
$\sigma$, $\tilde{\sigma}$, $S_{1/3}$ and $\tilde S_{-5/3}$ respectively;
The charged current interaction, with the $G^C$ coefficient, induces the 17
keV neutrino mode in $\beta$ decays. The existence of the additional neutral
current interactions in (\ref{fulll}) is implied by the broken \sutw{}
symmetry.

The extra, neutral current interactions in (\ref{fulll}) are extremely
important
for the study of the experimental constraints on $G^C$: The LEP experiments
\cite{LEP} put a lower bound of $O(50{\rm GeV})$ on all leptoquarks, that is,
leptoquark masses are of the order of the electroweak scale or higher. This
implies that $G^N_l$ and $G^N_\nu$ are expected to be similar in size and we
denote both by $G^N$. Similarly, we expect $\tilde G^N_l\sim \tilde G^N_\nu$
and denote both parameters by $\tilde G_N$. The coefficient $G^C$ is then
expected to be $\le \sqrt{G^N\tilde G^N}$. Therefore, bounds on
processes mediated by the neutral current interactions in (\ref{fulll}), will
be translated to bounds on the charged current coefficient $G^C$.

The most significant bound on $G^N$ arises from $eD$ scattering and atomic
Parity violation experiments. The parameters $C_{1d}$ and $C_{2d}$ (see
\cite{PDG}) are modified in our model to:
\begin{eqnarray}
C_{1d}&=&C_{1d}^{SM}+\frac{G^N_l}{G_F}\nonumber\\
C_{2d}&=&C_{2d}^{SM}-\frac{G^N_l}{G_F}\;. \label{C}
\end{eqnarray}
where the superscript $SM$ indicates standard model values. The experimental
value of $C_{1d}$ \cite{PDG} implies, for $m_t=100$ GeV, that
$G^N_l=(0.022\pm0.041)G_F$. (For $m_t=200$ GeV this range is slightly modified
to $G^N_l=(0.017\pm0.041)G_F$.) Looser bounds on $G^N_l$ can be extracted from
the experimental value of $C_{2u}-\frac{1}{2}C_{2d}$. If $\nu^x$ is $\nu^\tau$,
then very little is known about $\tilde G^N$, which could be as large as $G_F$
(or even larger), and therefore we cannot translate the results from $eD$
scattering and atomic Parity violation to a bound on $G^C$. However, if $\nu^x$
is $\nu^\mu$, measurements of $\nu^\mu N$ interactions imply that $\tilde
G^N_\nu=(-0.03\pm 0.02)G_F$\footnote{When discussing bounds arising from data
on $\nu N$ interactions, we use the results presented in
\cite{BPW}, which, in turn, where derived from \cite{Amaldi}}.
Combining the experimental results for $G^N$ and $\tilde G^N$ we find:
$G^N\tilde G^N=(-0.66\pm2.43)\cdot 10^{-3}G_F^2$. This value is
4.5$\sigma$ away from the minimum requested in our model: $G^N\tilde
G^N\ge0.01G_F^2$ (note that $G^N$ and $\tilde G^N$ should be positive
definite).  The lesson from studying the neutral current interactions of
the Lagrangian (\ref{fulll}) is therefore that $\nu^x$ is $\nu^\tau$
rather than $\nu^\mu$.

Next, we note the need for some $U(1)$ symmetries, to forbid unwanted
interactions of the $S$ and $\tilde S$ scalars (such symmetries are also
necessary for the model of ref. \cite{MPRT}): Without such symmetries,
the $S$ and $\tilde S$  leptoquarks may have many Yukawa interactions on top of
those of eq. (\ref{Yukawa}). The extra interactions will lead to various
unwanted effects, {\it e.g.}, FCNC processes in both the quark and lepton
sector. We therefore choose to conserve two combinations of lepton
numbers: $L_{e\tau}=L_e+L_\tau$ and  $L_\mu$
and also to impose an approximate $U(1)$  symmetry with the following charges:
\beq
Q(e_L,\nu^e_L)       = Q(e_R) = Q(d_R) = Q(\nu^i_R) = -1 ;,\label{Qn}
\eeq
\beq
Q(\tau_L,\nu^\tau_L) = Q(\tau_R) = Q(u_R) = 1 \;.\label{Qp}
\eeq
All other particles carry trivial $Q$ charges. $Q$ is free of QCD
anomalies.

Conservation of total lepton number $L=L_{e\tau}+L_\mu$ avoids neutrinoless
double $\beta$ decay. Also, since $\nu_\tau$ has nonvanishing mass, $L$
conservation means that RH neutrinos exist and the neutrinos are Dirac
particles. The $U(1)$ symmetries do not allow the RH neutrinos to participate
in the leptoquark interactions. Consequently, these neutrinos do not pose
any problem to primordial nucleosynthesis.

$L_{e\tau}$, $L_\mu$ and $Q$ conservation allow $S$ and $\tilde S$ the
Yukawa couplings of eq. (\ref{Yukawa}), but no others, thereby leading to
the effective Lagrangian (\ref{fulll}) which contains no FCNC terms. Note
however that $Q$ is only an approximate symmetry -- it is explicitly
broken by the masses of the up and down quarks and by neutrino
masses. Fortunately, this explicit breaking is extremely small, the
dimensionless parameter characterizing it being the Higgs Yukawa coupling
to the down quark: $y_d=e m_d/2 \sin\theta_W M_W\sim4\cdot 10^{-5}$. We
therefore find that the $S$ and $\tilde S$ leptoquarks may have extra
Yukawa couplings on top of those of (\ref{Yukawa}), but these couplings
({\it i}) must conserve $L_{e\tau}$ and $L_\mu$ and ({\it ii}) are
suppressed, relative to $g$ and $\tilde g$, by $y_\mu=4\cdot10^{-5}$. The
strong suppression ensures that the effects of the additional couplings be
unobservable in present and near future experiments.

We also note that the $Q$ symmetry is spontaneously broken at the QCD
scale when the up and down quarks condensate. However, the pseudo
Goldstone boson associated with this spontaneous breaking is the well
known $\pi^0$ and its interactions do not present any difficulty to our
model.

At this stage we are satisfied that the full four-Fermi interaction
induced by the $S$ and $\tilde S$ leptoquarks is given in (\ref{fulll}).
We also argued that, with the choice $\nu^x=\nu^\tau$, present
experimental data allows $G^N$ and $\tilde G^N$ to get large enough values
to be consistent with $G^C=0.1G_F$. The only difficulty still presented to the
model by experimental data is that such a large value for $G^C$ leads to
an unacceptable deviation from $e-\mu$ universality in $\pi$ decay. The
model's prediction for the ratio $R=\Gamma(\pi\longrightarrow
e\nu)/\Gamma(\pi\longrightarrow\mu\nu)$ is
\beq
R_{model}=R_{SM}(1+(\frac{G^C}{G_F})^2)=(1.246\pm0.001)\cdot 10^{-4} \;,
\label{Rth}
\eeq
where $R_{SM}=1.234\pm0.001$ \cite{Marciano} is the standard model prediction.
The latest experimental result \cite{Britton} is:
\beq
R_{exp}=(1.2265\pm0.0034(stat)\pm0.0044(sys))\cdot 10^{-4}\;.
\label{Rexp}
\eeq
Adding the statistical and systematic errors in quadrature, one
finds that the model's prediction for $R$ is 3.5$\sigma$ away from the
measured value.

This problem is typical to the ``unmixed 17 keV" scenario in general. Here
we suggest to solve it by assuming additional leptoquark interactions that
will enhance $\pi\longrightarrow\mu\nu$, thereby compensating  for the
enhancement of $\pi\longrightarrow e\nu$. We thus introduce a leptoquark $R$,
a singlet of \sutw{}, with 1/3 units of electromagnetic charge and
carrying $L_\mu=-1$ and $L_{e\tau}=Q=0$. The symmetries of the model allow $R$
to have the following Yukawa interactions:
\beq
\L^R_{Yukawa}=\sum_i h_i\left[(\nu^\mu_L)^tC^*(d^i_L)'
                               -(\mu_L)^tC^*u^i_L \right]R\;
\eeq
where $C$ is the charge conjugation matrix; $(u^i_L,(d^i_L)')$ is the
$i$'th generation quark doublet: $(d^i_L)'$ is the interaction eigenstate
which is related to the mass eigenstate $d^i_L$ by a
Cabibbo-Kobayashi-Maskawa rotation. Note that the $L_\mu$ and $L_{e\tau}$
charges of $R$ are different from those of $S$ and $\tilde S$.
Consequently, $R$ and $S_{1/3}$ cannot mix.

The effective Lagrangian induced by $R$ is:
\begin{eqnarray}
\L_{eff}^{R}&=&4\sum_{ij}H^N_{ij}
              \bar u^i_L\gamma^\mu u^j_L ~ \bar\mu_L\gamma_\mu\mu_L
\nonumber\\
            &+&4\sum_{ij}\tilde H^N_{ij}
              \bar d^i_L\gamma^\mu d^j_L ~ \bar\nu^\mu_L\gamma_\nu^\mu\mu_L
\nonumber\\
            &-&4\left[\sum_{ij}H^C_{ij}
              \bar u^i_L\gamma^\mu d^j_L ~ \bar\mu_L\gamma_\mu\nu^\mu_L
{}~ + ~ h.c. \right] \;,\label{moreL}
\end{eqnarray}
where we define a CKM rotated coupling $\tilde h_i=h_jV_{ji}$ and
\begin{eqnarray}
\frac{H^N_{ij}}{\sqrt2}&=&\frac{h_i^*h_j}{8M_L^2}
\nonumber\\
\frac{\tilde H^N_{ij}}{\sqrt2}&=&\frac{\tilde h_i^*\tilde h_j}{8M_L^2}
\nonumber\\
\frac{H^C_{ij}}{\sqrt2}&=&\frac{h_i^*\tilde h_j}{8M_L^2}
\;.\label{H}
\end{eqnarray}

We assume that there is a generation hierarchy for the $h_i$ couplings so
that $h_i\sim V_{i1}h_1$. Then, the $\tilde h_i$ couplings are also $\sim
V_{i1}h_1$ and $\tilde h_1=h_1(1+O(sin^2\theta_C))$. The hierarchy
assumption on the couplings has few important consequences: First, it
implies that all the effective terms in (\ref{moreL}) involving quarks of
the second and third generation, are strongly suppressed and their effects
are unobservable. Second, it means that, up to corrections of a few
percents (of order $\sin^2\theta_C$), $H_{11}^N$, $\tilde H_{11}^N$ and
$H_{11}^C$ are equal and we will denote them all by $H_{11}$. Third, one
finds that (up to corrections of a few percents) $H_{11}$ is real and
positive.

$H_{11}$ is the coefficient of the new contribution to $\pi
\longrightarrow\mu\nu$ decay. It adds up {\it coherently} to the standard
model amplitude, and being real and positive it will {\it enlarge} the
$\pi\longrightarrow\mu\nu$ width, thereby compensating for the enlarged
$\pi\longrightarrow e\nu$ width. Equation (\ref{Rth}) is modified to:
\beq
R_{model}=R_{SM}(1+(\frac{G^C}{G_F})^2-2\frac{H_{11}}{G_F})\;.\label{Rthmod}
\eeq
To keep $R_{model}$ within $2\sigma$ from $R_{exp}$ one has to request
that:
\beq
0.003<\frac{H_{11}}{G_F}<0.012\;. \label{rangeH}
\eeq
This range is consistent, within $1\frac{1}{4}\sigma$ with the range
allowed by $\nu^\mu N$ scattering experiments
$\frac{H_{11}}{G_F}=0.09\pm0.07$. We conclude that the $R$ leptoquark may
supply the necessary enhancement of $\pi\longrightarrow\mu\nu$ width, via
interactions that are consistent with all available experimental data.

Finally, we wish to provide the 17 keV neutrino with a decay mechanism that
will allow its lifetime to be shorter than $8.4\cdot10^{11}$ sec (see
\cite{GNP} and references therein). An interesting scenario was suggested in
\cite{BMR}, according to which the decay occurs in the pure RH sector.
Inspired by this idea, we suggest introducing a scalar $N$ carrying $(-2)$
units of total lepton number, $(+2)$ units of the $Q$ charge, and being a
singlet of $SU(2)_W\times U(1)_Y$. The $N$ scalar can couple only to pairs of
RH neutrinos:
\beq
\L^N_{Yukawa}=\sum_{ij}y_{ij}(\nu_R^i)^t C^*(\nu_R^j)N\label{N}
\eeq
Depending on its $L_{e\tau}$ and $L_\mu$ charges, $N$ may induce
$\nu_\tau\longrightarrow\bar\nu^\mu\nu^\mu\nu^e$ or
$\nu_\tau\longrightarrow\bar\nu^e\nu^e\nu^e$. We denote by $G_\nu$ the
effective four Fermi coefficient of these neutrino interactions. $G_\nu$ should
be larger than $5G_F$ to induce fast enough $\nu_\tau$ decay, and smaller than
$160 G_F$ to avoid the RH neutrinos from being in equilibrium with matter till
times that are forbidden by primordial nuclear synthesis \cite{BMR}.

Our model is now completely consistent with all experimental data and
cosmological requirements. We should mention though that there are
potential astrophysical problems: First, we did not venture to solve the
``solar neutrino problem" simultaneously with the 17 keV neutrino. If
solar neutrino experiments will prove that there is indeed a problem, the
model will have to be modified. Second, there is a bound \cite{SN} on the
mass of Dirac neutrinos with sterile RH component from the observation of
the neutrino signal of SN1987A \cite{snexp}. This bound is of the order of 10
keV but
may still allow for a 17 keV particle.

Summarizing, we suggested a model for the ``unmixed 17 keV neutrino", in which
this particle is the LH tau neutrino. The effective interaction responsible for
the kink observed in the $\beta$ decay spectra is induced in our model by
charge 2/3 leptoquarks. These leptoquark interactions also lead to an
enhancement by
$1\%$ of $\pi\longrightarrow e\nu$ decay relative to the standard model value.
To compensate for this enhancement we introduced another leptoquark, with
charge
1/3, whose interactions lead to enhancement of $\pi\longrightarrow \mu\nu$
decay. The ratio $R=\Gamma(\pi\longrightarrow e\nu)/\Gamma(\pi\longrightarrow
\mu\nu)$ can then be consistent with its experimentally measured value.
We also introduce scalars that induce the decay of the 17 keV
neutrino: The
neutrinos in our model have a RH component and are Dirac particles. The scalars
that induce the decay interact only with the RH neutrinos, so their couplings
can be large relative to $G_F$ and consequently the decay rate is fast enough
to obey cosmological constraints. We had to impose some global $U(1)$
symmetries, to avoid troublesome Yukawa couplings of our leptoquarks. Such
symmetries must be imposed in every unmixed 17 keV neutrino model. Here we
chose to
conserve $L_{e\tau}=L_e+L_\tau$, $L_\mu$ and an approximate U(1) symmetry
whose charges where defined above. Our model is completely consistent with all
experimental data and cosmological constraints, it does however have potential
astrophysical problems: It does not contain an explanation for the ``solar
neutrino problem", in case this will indeed prove to be a problem. Also,
analyses of the SN1987A neutrino pulse imply that the masses of Dirac neutrinos
should not exceed $O(10){\rm keV}$. A 17 keV Dirac neutrino may be excluded in
the future if this bound is significantly improved.

\vskip 2.5cm
\noindent
{\bf Acknowledgements}\\
I thank Shmuel Nussinov for drawing my attention to ref. \cite{MPRT}.
I thank him and especially Neil Marcus for helpful discussions and comments.

\end{document}